\documentstyle[aps,preprint,eqsecnum,prd]{revtex}

\begin{document}

\draft

\title{Testing QCD Sum Rule techniques on the lattice}

\author{Derek B. Leinweber\cite{presaddr}}

\address{
Department of Physics, The Ohio State University \\
174 West 18'th Avenue, Columbus, OH 43210--1106 \\
and \\
Department of Physics, University of Maryland, College Park, MD 20742
}

\date{March 1994}

\maketitle

\begin{abstract}
Results for the first test of the ``crude'' QCD continuum model,
commonly used in QCD Sum Rule analyses, are presented for baryon
correlation functions.  The QCD continuum model is found to
effectively account for excited state contributions to the short-time
regime of two-point correlation functions and allows the isolation of
ground state properties.  Confusion in the literature surrounding the
physics represented in point-to-point correlation functions is also
addressed.  These results justify the use of the ``crude'' QCD
continuum model and lend credence to the results of rigorous QCD Sum
Rule analyses.
\end{abstract}


\narrowtext

\section{INTRODUCTION}
\label{intro}

   Point-to-point correlation functions have been the subject of
intense study since the early days of lattice field theory.  In the
quest for an {\it ab initio} determination of the low-lying hadron
spectrum, lattice QCD investigations have focused on the large
Euclidean-time tails of three-momentum-projected two-point correlation
functions.  As such, the short time regime, where excited state
contributions are significant, has simply been discarded.

   The lattice approach to QCD allows the determination of correlation
functions deep in the nonperturbative regime.  In contrast, the QCD
Sum Rule (QCD-SR) method \cite{shifman79} is restricted to the near
perturbative regime of the truncated Operator Product Expansion (OPE).
In this regime, one cannot ignore the contributions of excited states
in a phenomenological description of QCD correlation functions.  To
account for these contributions, the so-called ``crude'' QCD continuum
model is introduced \cite{shifman79}.  This model exploits the leading
terms of the OPE, and introduces a sharp threshold marking the onset
of the QCD continuum.  The contributions of this model relative to the
ground state, whose properties one is really trying to determine, are
not small.  They are typically 10 to 50\% \cite{leinweber90}.  The
validity of this model is relied upon to cleanly remove the excited
state contaminations.

   This investigation examines the physics in the near perturbative
regime of point-to-point correlation functions where QCD-SR analyses
are performed.  The QCD continuum model is constructed for
three-momentum-projected Euclidean-time two-point functions following
the techniques established in the QCD-SR approach.  This model will
then be used as a probe of the physics represented in lattice QCD
correlation functions, and as a test of QCD-SR techniques.  This
investigation explores some of the ideas briefly summarized in Ref.\
\cite{leinweber94e} in greater depth and detail.

   Some attention has recently been given to the behavior of lattice
point-to-point correlation functions in the near perturbative regime
\cite{chu93a,chu93b}.  There the focus was on correlation functions in
coordinate space.  It was concluded that the QCD-SR-inspired continuum
model was sufficient to describe the lattice correlation functions
over the calculated range.  Using the entire lattice correlation
function (including the deep nonperturbative regime), ground state
masses were extracted and were found to agree with conventional
lattice analyses.

   Here the emphasis is on determining ground state properties by
examining only the first few points of the lattice correlation
function in the near perturbative regime.  This is similar in spirit
to QCD-SR analyses.  In this manner, the validity of the QCD continuum
model is rigorously tested.  By extending the analysis interval of the
correlation function deeper into the nonperturbative regime, the
evolution of fitted ground state properties may be monitored.  A
sensitivity to the analysis interval in the extracted parameters would
indicate a failure of the QCD-SR-inspired continuum model.  Comparisons
are made with conventional lattice results where ground state
properties are simply extracted from the tails of the correlation
functions.  The importance of higher order terms of the OPE in the
formulation of the QCD continuum model is also examined.  These terms
were not investigated in Ref.\ \cite{chu93a,chu93b}.

   One would also like to evaluate whether these techniques are useful
in analyzing other lattice QCD correlation functions.  Such techniques
may be of practical use for analyzing two-point correlation functions
which become noisy prior to a clear ground state domination, such as
heavy-light meson correlators.  In a subsequent paper
\cite{leinweber94b}, these techniques will be exploited to investigate
nucleon properties obtained from unconventional interpolating fields.

   The outline of this paper is as follows.  Section \ref{lattcorrfun}
introduces the definition of the two-point function examined in this
investigation and summarizes the lattice techniques and parameters.
In Section \ref{contmodel}, the QCD continuum model is derived for
three-momentum-projected two-point correlation functions in Euclidean
space.  Issues associated with relating the lattice and continuum
(lattice spacing $a \to 0$) formulations of the model are discussed
here.  Fits of the nucleon correlator are presented in Section
\ref{corrfits}.  The validity of the QCD continuum model as
implemented in the QCD Sum Rule approach is evaluated in Section
\ref{SRtest}.  The fit parameters are compared with other approaches
in Section \ref{compappr}.  In section \ref{physint}, the physics
represented in two-point correlation functions is discussed, where
erroneous conclusions in the literature are addressed.  Finally,
Section \ref{concl} summarizes the findings of this investigation.

\section{LATTICE CORRELATION FUNCTIONS}
\label{lattcorrfun}

\subsection{Interpolating Fields}

   In lattice calculations, the commonly used interpolating field for
the proton has the form
\begin{equation}
\chi_1(x) = \epsilon^{abc}
                 \left ( u^{Ta}(x) C \gamma_5 d^b(x) \right ) u^c(x)
\, .
\label{chiN1}
\end{equation}
Here, we follow the notation of Sakurai \cite{sakurai67}.  The Dirac
gamma matrices are Hermitian and satisfy $\left \{ \gamma_\mu ,
\gamma_\nu \right \} = 2 \, \delta_{\mu \nu}$, with $\sigma_{\mu \nu}
= {1 \over 2i} \left [ \gamma_\mu , \gamma_\nu \right ] $.  $C =
\gamma_4 \gamma_2$ is the charge conjugation matrix, $a,\ b,\ c$ are
color indices, $u(x)$ is a $u$-quark field, and the superscript $T$
denotes transpose.  Dirac indices have been suppressed.

   In the QCD Sum Rule approach, it is common to find linear combinations
of this interpolating field and
\begin{equation}
\chi_2(x) = \epsilon^{abc}
                 \left ( u^{Ta}(x) C d^b(x) \right ) \gamma_5 u^c(x)
\, ,
\label{chiN2}
\end{equation}
which vanishes in the nonrelativistic limit.  With the use of
Fierz relations, the combination of the above two interpolating fields
with a relative minus sign may be written
\begin{eqnarray}
\chi_{\rm SR}(x) &=& \epsilon^{abc}
                 \left ( u^{Ta}(x) C \gamma_\mu u^b(x) \right )
                 \gamma_5 \gamma^\mu d^c(x) \, , \nonumber \\
           &=& 2 \left ( \chi_2 - \chi_1 \right ) \, ,
\label{chiSR}
\end{eqnarray}
giving the proton interpolating field often found in QCD Sum Rule
calculations \cite{leinweber90,ioffe81,ioffe83}.

   Here the task is to evaluate the validity of the QCD-SR-inspired
continuum model.  This issue is best addressed with the use of the
interpolating field $\chi_1$ of (\ref{chiN1}).  For comparison with
other calculations, results for $\chi_{\rm SR}$ are also reported.

   The $\Delta$ resonance is also considered for comparison with the
nucleon results.  For simplicity, the $\Delta^{++}$ interpolating
field is used,
\begin{equation}
\chi_\Delta^\mu(x) = \epsilon^{abc}
                 \left ( u^{Ta}(x) C \gamma^\mu u^b(x) \right ) u^c(x)
\, .
\label{chiDelta}
\end{equation}

\subsection{Two-Point Function}

   Two-point correlation functions are used to determine hadron
masses.  Consider the following two-point function for the nucleon,
\begin{equation}
G_2(t,\vec p) = \sum_{\vec x}\, e^{-i \vec p \cdot \vec x}
{\rm tr} \Bigl [ \Gamma_4
\bigm < 0 \bigm | T \{ \chi_1 (x)\, \overline \chi_1 (0) \} \bigm |
0 \bigm > \Bigr ]  \, .
\label{twopoint}
\end{equation}
Here, $\Gamma_4 = (1 + \gamma_4)/4$ projects positive parity states
for $\vec p = 0$, and tr indicates the trace over Dirac indices.
Correlation functions at the quark level are obtained through the
standard procedure of contracting out time-ordered pairs of quark
field operators,
\widetext
\begin{eqnarray}
G_2(t,\vec p) = \sum_{\vec x} e^{-i \vec p \cdot \vec x}
   tr \Biggl [ \Gamma_4 \, \epsilon^{abc} \epsilon^{a'b'c'}
&& \biggl \{
   S_{u}^{a a'}(x,0) \, tr \left [ S_{u}^{b b'}(x,0) \,
   \widetilde C S_d^{c c' \, T}(x,0) \widetilde C^{-1}
   \right ] \nonumber \\
&& \quad + S_{u}^{a a'}(x,0) \, \widetilde C S_d^{c c' \, T}(x,0)
   \widetilde C^{-1} \, S_{u}^{b b'}(x,0) \biggr \} \Biggr ] \, ,
\label{twoptquark}
\end{eqnarray}
\narrowtext
where $\widetilde C = C \gamma_5$, and $S_u^{a a'}(x,0) = T \left \{
u^a(x), \, \overline u^{a'}(0) \right \}$, {\it etc}.

\subsection{Lattice Techniques}

   Here we briefly summarize the lattice techniques used to calculate
(\ref{twoptquark}).  Additional details may be found in Ref.\
\cite{leinweber91}.  Wilson's formulation is used for both the gauge
and fermionic action.  $SU(2)$-isospin symmetry is enforced by
equating the Wilson hopping parameters $\kappa_u = \kappa_d = \kappa$.
Three values of $\kappa$ are selected and are denoted $\kappa_1 =
0.152$, $\kappa_2 = 0.154$ and $\kappa_3 = 0.156$.  To make contact
with the physical world, the mass, continuum threshold and
interpolating field coupling strengths calculated at the three values
of $\kappa$ are linearly extrapolated to $\kappa_{\rm cr}=0.159\,8(2)$
where an extrapolation of the squared pion mass vanishes.  Differences
between linear extrapolations to $m_\pi=0$ as opposed to the physical
pion mass are small and are neglected in the following.

   Twenty-eight quenched gauge configurations were generated
\cite{leinweber91} by the Cabibbo-Marinari \cite{cabibbo82}
pseudo-heat-bath method on a $24 \times 12 \times 12 \times 24$
periodic lattice at $\beta=5.9$. Configurations were selected after
5000 thermalization sweeps from a cold start, and every 1000 sweeps
thereafter \cite{correlations}.

   Dirichlet boundary conditions are used for fermions in the time
direction.  Time slices are labeled from 1 to 24, with the
$\delta$-function source at $t=4$.  To minimize noise in the Green
functions, the parity symmetry of the correlation functions, and the
equal weighting of $\{U\}$ and $\{U^*\}$ gauge configurations in the
lattice action are exploited.  The nucleon mass determined from
$\chi_1$ of (\ref{chiN1}) is used to set the lattice spacing.  This
estimate lies between other estimates based on the string tension or
the $\rho$-meson mass.  The lattice spacing is determined to be
$a=0.132(4)$ fm, and $a^{-1} = 1.49(5)$ GeV.

   Statistical uncertainties in the lattice correlation functions are
estimated by a single elimination jackknife \cite{efron79}.  A
covariance matrix fit of the pole plus QCD continuum model over a
typical range of 15 time slices is likely to be unreliable for 28
gauge configurations \cite{michael94}.  Instead we use standard
statistical error analysis in which correlations among the fit
parameters are accounted for.  The Gauss-Newton method is used to
minimize $\chi^2$.  Uncertainties are taken from the standard error
ellipse \cite{rpp92} at $\chi^2 = \chi^2_{\rm min} + 1$.

\subsection{Mean-Field Improvement}

   In this analysis, the nucleon coupling strength (or residue of the
nucleon pole) is determined in absolute terms, without resorting to a
ratio of the QCD continuum contributions as done in
\cite{chu93a,chu93b}.  Our approach requires the use of mean-field
improvement \cite{lepage92,lepage93}
\begin{equation}
\psi^{\rm Continuum} \equiv { \sqrt{2 \kappa} \over a^{3/2} } \,
                            \psi^{\rm Lattice}  \, ,
\end{equation}
where
\begin{equation}
\sqrt{2 \kappa} \to
\left ( 1 - { 3 \kappa \over 4 \kappa_{\rm cr}} \right )^{1/2} \, ,
\label{mfi}
\end{equation}
to account for otherwise large renormalization factors.  While
(\ref{mfi}) is derived for heavy quarks \cite{lepage92}, its use in
light quark studies has already been suggested in \cite{lepage93}.
The $\kappa$ dependence of these two wave function normalizations is
very different and use of the mean-field improved normalization is
crucial to recovering the correct mass independence of the Wilson
coefficient of the identity operator.  This point will be further
illustrated in the following section.

   The implementation of Wilson fermions on the lattice induces mixing
between the composite nucleon interpolating fields of (\ref{chiN1})
and (\ref{chiN2}), reflecting the breaking of chiral symmetry
\cite{richards87}.  This mixing is discussed in more detail in Ref.\
\cite{leinweber94b}.  The important point is that the mixing of the
interpolating fields $\chi_1$ and $\chi_2$ is negligible.  As a
result, it is possible to identify the properties of these
interpolating fields determined on the lattice with those of their
continuum $(a \to 0)$ counterparts to a good approximation.  The
principle renormalization constant $C_1^{\rm L}$ has been determined
in the mean-field approach \cite{lepage93} and is used in the
following.  The renormalization at the scale of $1/a$ is
\begin{equation}
\chi^{\rm Continuum} = {Z_{\chi_N} \over a^{9/2}} \,
                       \chi^{\rm Lattice} \,
\left ( 1 - { 3 \kappa \over 4 \kappa_{\rm cr}} \right )^{3/2} \, ,
\end{equation}
and $Z_{\chi_N} = (1 - 0.73 \, \alpha_V) \simeq 0.80$ at $\beta =
5.9$.
%

\section{THE EUCLIDEAN SPACE FORMULATION OF THE QCD CONTINUUM MODEL}
\label{contmodel}

\subsection{Spectral Representation}

   At the phenomenological level, the two-point function is calculated
by inserting a complete set of states $N^i$ between the interpolators
of (\ref{twopoint}) and defining
\begin{equation}
\bigm < 0 \bigm | \chi_1 (0) \bigm | N^i,p,s \bigm > \, =
\lambda_1^i\, u(p,s) \, .
\end{equation}
Here, the coupling strength, $\lambda_1^i$, measures the ability of
the interpolating field $\chi_1$ to annihilate the i'th nucleon
excitation.  For $\vec p = 0$ and Euclidean time $t \to \infty$, the
ground state dominates and
\begin{equation}
G_2(t) \to  \lambda_1^2 e^{-M_N t} \, .
\end{equation}
We note here that the exponential suppression of excited states is
somewhat similar to that encountered in Borel improved QCD-SR analyses.
There the square of the mass appears in the exponential as $\exp(-
M_N^2 \, \tau^2)$ where $\tau$ is the inverse Borel parameter.  The
spectral representation is defined by
\begin{equation}
G_2(t) = \int_0^\infty \, \rho(s) \, e^{-st} \, ds \, ,
\label{SpectralRepr}
\end{equation}
and the spectral density is
\begin{equation}
\rho(s) = \lambda_1^2 \, \delta(s-M_N) + \zeta(s) \, ,
\end{equation}
where $\zeta(s)$ provides the excited state contributions.

   While it may be tempting to fit the correlation function in the
shorter time regime by including additional poles in the spectral
density,
\begin{equation}
\rho(s) = \lambda_N^2 \, \delta(s-M_N) +
          \lambda_{N^*}^2 \, \delta(s-M_{N^*}) +
          \cdots \, ,
\label{TwoPoles}
\end{equation}
such an approach fails in a number of ways.  If one simply includes a
couple of poles for the spectral density, one finds that the mass of
the excitation is completely determined by the leading time slice that
is considered in the fit.  This is illustrated in Figure
\ref{twopolefit} where a fit of time slices 7 through 20 is shown.  If
time slice 6 is included, the mass of the second pole increases to
accommodate the earlier time slice with smaller uncertainties.

   One also finds that the coupling strength $\lambda_{N^*}$ exceeds
$\lambda_N$ by an order of magnitude.  Our physical intuition
suggests the opposite.  The coupling strength should decrease for
excited states as these states are expected to have broader wave
functions.  The probability of finding three quarks at the same space
time point is therefore smaller.  In addition, such an approach is a
poor representation of the physics believed to be there as it neglects
the natural widths of the resonances and the QCD continuum of multiple
hadron states.  The correlation function is probably best described by
many states of diminishing coupling strengths and increasing widths.
It may be that the QCD-SR-inspired continuum model is an efficient way
of characterizing this physics.

\subsection{ Operator Product Expansion (OPE)}

   The form of the spectral density used in the QCD continuum model is
determined by the leading terms of the OPE surviving in the limit $t
\to 0$.  The calculation of these terms proceeds in the usual fashion
of contracting out time-ordered pairs of quark field operators.  For
the nucleon interpolating field of (\ref{chiN1}) one has
\widetext
\begin{equation}
G_2(t,\vec p=0) = \sum_{\vec x}\, {\rm tr} \biggl [ \, \Gamma_4 \,
\epsilon^{abc} \epsilon^{a'b'c'}
\biggl \{
S_u^{a a'} \, {\rm tr} \left ( S_u^{b b'} \, \gamma_5 C \, S_d^{T
c c'} C^{-1} \gamma_5 \right )
+ S_u^{a a'} \, \gamma_5 C \, S_d^{T b b'} C^{-1} \gamma_5 \,
  S_u^{c c'} \biggr \} \; \biggr ]  \, . \nonumber
\end{equation}
\narrowtext
In Euclidean space and coordinate gauge the quark propagator
has the expansion
\begin{eqnarray}
S_q^{a a'} &=& {1 \over 2 \pi^2} \, {\gamma \cdot x \over x^4}
               \delta^{a a'}
           + {1 \over \left ( 2 \pi \right )^2} \, { m_q \over x^2}
             \delta^{a a'} \nonumber \\
         &&- {1 \over 2^2 3} \bigm < : \overline q q : \bigm >
             \delta^{a a'}
           + \cdots \, ,
\end{eqnarray}
and $G_2(t)$ has the following OPE
\begin{eqnarray}
G_2(t) &\simeq& {3 \cdot 5^2 \over 2^8 \pi^4} \, \biggl ( \,
{1 \over t^6} + {28 \over 25} \, {m_q a \over t^5}
              + {14 \over 25} \, {m_q^2 a^2 \over t^4} \nonumber \\
            &&- {56 \pi^2 \over 75} \, {\bigm < : \overline q q :
                                        \bigm > a^3
                \over t^3}
              + \cdots \biggr ) \, .
\label{OPE}
\end{eqnarray}
The quark mass used in the OPE is obtained from the standard relation
\begin{equation}
m_q \, a = {1 \over 2}
  \left ( {1 \over \kappa} - {1 \over \kappa_{\rm cr}} \right ) \, ,
\end{equation}
and is renormalized to one loop \cite{daniel92,mfimq}.

\subsection{QCD Continuum Model Contributions}

   The spectral density used in the QCD continuum model is defined by
equating (\ref{SpectralRepr}) and (\ref{OPE}).  Thus, $\rho(s)$ is
given by the following Laplace transform pairs
\begin{equation}
{1 \over t^n} \to {1 \over \left ( n-1 \right )!}\, s^{n-1} \, ,
\qquad \begin{array}{l} n = 1, 2, 3, \ldots \\ {\rm Re}\ t > 0
\end{array} \, .
\end{equation}
The QCD continuum model is defined through the introduction of a
threshold which marks the effective onset of excited states in the
spectral density.  This model is motivated by duality arguments as
realized in the $Q^2$ dependence of the experimental cross section for
$e^+ e^- \to $ hadrons.  The QCD continuum model contribution is
\begin{equation}
\int_{s_0}^\infty \, \rho(s) \, e^{-st} \, ds
= e^{-s_0 t} \, \sum_{n=1}^6 \, \sum_{k=0}^{n-1} \: {1 \over k!} \,
{s_0^{k} \over t^{n-k} } \: C_n \, {\cal O}_n  \, ,
\end{equation}
where $C_n$ and ${\cal O}_n$ are the Wilson coefficient and normal
ordered operator of the term $t^{-n}$ in (\ref{OPE}).  The
phenomenology of $G_2(t)$ is summarized by
\begin{mathletters}
\label{phenom}
\begin{eqnarray}
G_2(t) &=& \lambda_1^2 \, e^{-M_N t} + \xi \,
\int_{s_0}^\infty \, \rho(s) \, e^{-st} \, ds  \, ,
\label{phenomCrypt} \\
&=& \lambda_1^2 \, e^{-M_N t} + \xi \,
    {3 \cdot 5^2 \over (2^8 \pi^4)} \,
e^{-s_0 t} \, \Biggl (  \label{phenomExpl} \\
&& \quad \left \{ {1 \over t^6} + {s_0 \over t^5}
+ {1 \over 2}\,{s_0^2 \over t^4} + {1 \over 6}\,{s_0^3 \over t^3}
+ {1 \over 24}\,{s_0^4 \over t^2} + {1 \over 120}\,{s_0^5 \over t}
\right \} \nonumber \\
&& \quad + {28 \, m_q a \over 25} \,  \left \{
  {1 \over t^5} + {s_0 \over t^4}
+ {1 \over 2}\,{s_0^2 \over t^3} + {1 \over 6}\,{s_0^3 \over t^2}
+ {1 \over 24}\,{s_0^4 \over t}
\right \} \nonumber \\
&& \quad + {14 \, m_q^2 a^2 \over 25} \,  \left \{
  {1 \over t^4} + {s_0 \over t^3}
+ {1 \over 2}\,{s_0^2 \over t^2} + {1 \over 6}\,{s_0^3 \over t}
\right \} \nonumber \\
&& \quad - {56 \, \pi^2 \bigm < \overline q q \bigm > a^3 \over 75} \,
\left \{
  {1 \over t^3} + {s_0 \over t^2}
+ {1 \over 2}\,{s_0^2 \over t}
\right \} + \cdots \Biggr ) \, .  \nonumber
\end{eqnarray}
\end{mathletters}
Here we have introduced an additional parameter, $\xi$, governing the
strength of the QCD continuum model.  Strictly speaking, $\xi = 1$ in
the continuum limit, $a \to 0$, but here is optimized with
$\lambda_1$, $M_N$, and $s_0$ to account for enhancement of the
correlator in the short time regime due to lattice anisotropy.  Its
deviation from 1 is a reflection of the fact that we are matching the
model formulated in the continuum limit to the lattice results at
finite $a$.  Ref.\ \cite{chu93b} found the anisotropy to be large for
$x-x_0 < 6$ for free quark correlators and remain large in their
interacting simulation at $\beta = 5.7$.  At $\beta = 5.9$ there is
some hope that anisotropy issues will be less problematic for the
Fourier transformed correlators presented here.  However, at very
short times the quarks are essentially free and the anisotropy must be
accommodated.  With this approach, the effects of lattice anisotropy
are absorbed through a combination of a larger QCD continuum model
strength $(\xi > 1)$ and marginally larger continuum threshold
$(s_0)$.

\subsection{Lattice Cutoffs}

   Unlike coordinate space lattice analyses \cite{chu93a,chu93b},
infrared lattice artifacts are not a significant problem for this
approach.  The Fourier transform weight $\exp(-i \vec p \cdot \vec x)$
is correct for all propagator paths including those which wrap around
the lattice spatial dimensions.

   The ultraviolet lattice cutoff may be modeled in a manner similar
to that for the QCD continuum model.  The upper limit of the integral
over the spectral density of (\ref{phenomCrypt}) is cut off at
$\Lambda$, and additional terms appear in (\ref{phenomExpl}).  As an
example, consider the leading term of the OPE.  In this case,
\begin{equation}
{1 \over t^6} = \int_0^\infty \, {1 \over 5!} \, s^5 \, e^{-st} \, ds
\, ,
\end{equation}
is modified to
\widetext
\begin{equation}
\int_0^\Lambda \, {1 \over 5!} \, s^5 \, e^{-st} \, ds =
{1 \over t^6} - e^{-\Lambda t} \,
  \left \{ {1 \over t^6} + {\Lambda \over t^5}
+ {1 \over 2}\,{\Lambda^2 \over t^4}
+ {1 \over 6}\,{\Lambda^3 \over t^3}
+ {1 \over 24}\,{\Lambda^4 \over t^2}
+ {1 \over 120}\,{\Lambda^5 \over t}
\right \} \, .
\label{uvcutoff}
\end{equation}
\narrowtext
Figure \ref{cutoffcomp} illustrates the $C_6/t^6$ term with the
ultraviolet cutoff correction (dashed line) fit to the lattice data
{}from $t=5 \to 7$ by optimizing $\Lambda$ and $C_6$.  The
corresponding curve for $C_6/t^6$ without the ultraviolet correction
(solid line) is also indicated.  The optimum value for $\Lambda$ is
4.55(3) which is not too different from the momentum cutoff of $\pi$.
Of course, the two numbers need not agree since we are modeling the
ultraviolet cutoff and matching a continuum formulation, $(a \to 0)$,
to the lattice.  The correction term accounting for the ultraviolet
lattice cutoff is negligible by $t-t_0=2$.  Rather than introducing an
extra parameter in the fit, the correction is simply neglected, and
all fits begin at $t=6$ in the following.

   A similar modification of the correlator due to the lattice
regularization was not accounted for in Ref.\ \cite{chu93b}, when
comparing hadron correlators calculated on the lattice with free quark
propagators, and correlators using continuum, $(a \to 0)$,
propagators.  Inclusion of these effects in Fig. 3 of Ref.
\cite{chu93b} would suppress the continuum curve at short distances
such that it may continue to follow the diagonal elements of the
lattice data more closely.

\subsection{QCD Continuum Model Summary}

   Figure \ref{ContModelCont} displays the total and individual
contributions to the QCD continuum model term of (\ref{phenom}).  For
comparison, the total and individual contributions to the OPE of
(\ref{OPE}) are illustrated in Figure \ref{OPEcont}.  The higher order
terms of the OPE become significant at large time separations.
Therefore, a truncated OPE will have a corresponding upper limit in
time separations within which reasonable convergence is maintained.
In contrast, the exponential of the continuum threshold in the QCD
continuum model acts to suppress the higher order terms.  The relative
contributions of the subsequent terms of the OPE to the QCD continuum
model are well ordered throughout the fitting regime, and inclusion of
the first few terms of the OPE is adequate.  Indeed, the contributions
{}from the $m_q^2$ and quark-condensate terms make negligible
contributions relative to the identity and $m_q$ terms.  These
higher-order terms are not included in the following analysis.  In
summary, the phenomenological description of the correlation function
is taken to be
\begin{eqnarray}
G_2(t) &=& \lambda_1^2 \, e^{-M_N t} + \xi \,
    {3 \cdot 5^2 \over (2^8 \pi^4)} \,
e^{-s_0 t} \, \Biggl (  \label{phenomTrun} \\
&& \quad \left \{ {1 \over t^6} + {s_0 \over t^5}
+ {1 \over 2}\,{s_0^2 \over t^4} + {1 \over 6}\,{s_0^3 \over t^3}
+ {1 \over 24}\,{s_0^4 \over t^2} + {1 \over 120}\,{s_0^5 \over t}
\right \} \nonumber \\
&& \quad + {28 \, m_q a \over 25} \,  \left \{
  {1 \over t^5} + {s_0 \over t^4}
+ {1 \over 2}\,{s_0^2 \over t^3} + {1 \over 6}\,{s_0^3 \over t^2}
+ {1 \over 24}\,{s_0^4 \over t}
\right \} \Biggr ) \, .  \nonumber
\end{eqnarray}

   The shaded areas of Figure \ref{OPEcont} indicate the regions
typically excluded form Borel-transformed QCD-SR analyses.  The
remaining region is far from the regime where the ground state pole
dominates the correlation function.  A rigorous uncertainty analysis
is needed to ascertain the predictive power of QCD Sum Rules
\cite{leinweber94c}.

\section{NUCLEON CORRELATOR FITS}
\label{corrfits}

   Among the first of things to verify is the leading time dependence
of $G_2(t)$.  Here we consider the lattice data at $\kappa = 0.156$
where $m_q$ corrections are smallest and one has the best chance to
verify that the dependence is indeed proportional to $1/t^6$.  Fitting
time slices $t=6$ and $t=7$ to the functional forms of $t^{-7}$,
$t^{-6}$, and $t^{-5}$ yields a $\chi^2/$dof of 80, 4.8, and 230
respectively, suggesting the predominant behavior is $1 / t^6$ as
anticipated.

   Our purpose is to test whether the nucleon mass and coupling
strength can be obtained accurately from a fit considering only the
first few points of the correlation function.  The lattice correlation
functions \cite{leinweber91} are fit with (\ref{phenomTrun}) in a
four-parameter search of $\lambda_1$, $M_N$, $s_0$ and $\xi$ in
analysis intervals from $t=6 \to t_f$ where $t_f$ ranges from 11
through 23.  Figure \ref{NuclCorrFn} illustrates these 13 fits of the
lattice correlation function at the smallest value of $\kappa$
considered.  The statistical uncertainties are smallest for this quark
mass and provides the best opportunity for revealing structure in the
lattice correlation function that is not accounted for by the QCD
continuum model. The typical $\chi^2/$dof for these pole plus QCD
continuum model fits is 0.6.

   The nucleon mass determined in each of the intervals is plotted as
a function of $t_f$ in Figure \ref{Nmass}.  Pole plus QCD continuum
model fits are compared with simple pole fits.  Figure \ref{Nlambda}
illustrates similar results for the coupling strength.  It is
interesting to see that the simple pole determination of $\lambda_1$
fails to form a plateau at large time separations.  Similar results
are seen for the larger values of $\kappa = 0.154$ and 0.156.  The fit
parameters extracted for the region $t=6 \to 20$ are summarized in
Table \ref{Tab:chi11C5ope} \cite{anomdim}.  The plateau in $M_N$ and
$\lambda_1$ for pole plus QCD continuum model fits indicates that the
QCD continuum model effectively accounts for excited state
contaminations in the correlation functions.  The search parameters
$s_0$ and $\xi$ display a similar plateau as one might expect, since
the regime in which these parameters are largely determined is common
to all intervals.

   Table \ref{Tab:chi11C5ope} also indicates a value for $\xi$
obtained by fitting $\xi \times$ the first two terms of the OPE to
time slices $t=6$ and $7$.  Higher order terms not included in
(\ref{OPE}) are expected to be significant and therefore $\xi$
obtained in this manner is only a rough estimate.  Its similarity to
that obtained from the QCD continuum model verifies the deviation of
$\xi$ from unity is a lattice artifact and not a failure of the QCD
continuum model.  It should also be mentioned that $\xi$ has been
defined with respect to the Wilson coefficient of the identity
operator only to leading order in perturbation theory.  $\alpha_s$ and
leading log corrections are expected to reduce $\xi$ by about 25\%.

   A similar fit including the Wilson coefficient of the $m_q$
correction, $C_5$, as a search parameter confirms the OPE value for
the coefficient ratio $C_5/C_6$.  The fact that this estimate
agrees with the OPE prediction confirms the role of $\xi$ as
an overall continuum model strength.  As such, $\xi$ is expected to be
independent of the quark mass in the same manner that $C_6$, the
coefficient of the identity operator, is.  To demonstrate that this is
in fact the case we present Figure \ref{xiMassDep}.  Here the quark
mass dependence of $\xi$ is illustrated (solid curve) for the three
values of $\kappa$ considered on the lattice as well as the value
linearly extrapolated to $\kappa_{\rm cr}$ where the pion mass
vanishes.  $\xi$ is nearly independent of the quark mass.  This
behavior is also in accord with the fact that enhancement of the
correlator due to lattice anisotropy is largest at very short time
separations where the OPE is dominated by the identity operator.

   This result is not trivial, as it requires the use of the OPE value
for $C_5$ and the use of mean-field-improved operators.  If the
na\"{\i}ve wave function normalization, $\sqrt{2 \kappa}$, is used,
$\xi$ varies as indicated by the dashed line plotted in Figure
\ref{xiMassDep}.

   In summary, the pole plus QCD continuum model allows the extraction
of nucleon ground state properties, even for an interval as small as
$t = 6 \to 11$.  The techniques examined here may be useful in
analyzing correlation functions that suffer a loss of signal prior to
clear ground state domination.  These techniques are used in a
subsequent paper investigating nucleon properties obtained from
unconventional interpolating fields \cite{leinweber94b}.

\section{QCD SUM RULE TEST}
\label{SRtest}

   In the sum rule approach $\xi$ is fixed to unity by the OPE.  To
test the QCD-SR method as closely as possible to its actual
implementation, we will use the value for $\xi$ obtained in the
previous fit from $t = 6 \to 20$.  This value is similar to that
obtained by fitting (\ref{OPE}) to the first few time slices of the
lattice data.

   Figure \ref{NcorrelSRT} illustrates the correlation function fits
for the lightest quark mass at $\kappa=0.156$ for the three-parameter
search of $\lambda_1$, $M_N$ and $s_0$, in analysis intervals from
$t=6 \to t_f$ where $t_f$ ranges {}from 9 through 23.  While ideally
we would like to start at $t_f=7$, a minimum of 4 points is required
for the fit with uncertainty estimates for the fit parameters.  The
typical $\chi^2/$dof for these pole plus QCD continuum fits is 0.53.
For the heaviest quark mass where the statistical uncertainties are
smallest the typical $\chi^2/$dof is 0.6.

   Using a pole plus QCD continuum model, one can extract ground state
nucleon properties using only the first few points of the correlation
function.  The stability of the nucleon mass to the analysis interval
is displayed in Figure \ref{NmassSRT}.  The nucleon coupling strength
is illustrated in Figure \ref{NlambdaSRT} where it is plotted as a
function of the analysis interval.  All three fit parameters are
stable as functions of $t_f$.  Moreover, the nucleon mass and coupling
strength agree with the values extracted from the tail of the
correlation function with a simple pole fit from $t_f - 7 \to t_f$
when $t_f \sim 20$ \cite{boundary}.  This indicates the QCD continuum
model inspired by QCD Sum Rules is successful in quantitatively
accounting for excited states in point-to-point correlation functions.
However, interplay between the fit parameters gives rise to rather
large uncertainties at $t_f = 9$.

   The smallest analysis interval considered here is still somewhat
generous relative to the interval considered in QCD-SR analyses.  Borel
improved Sum Rules including terms to dimension 8 in the OPE are
typically analyzed in the regime $\tau = 1/ M_{\rm Borel} = 0.15 \to
0.35$ fm corresponding to $t \simeq 5 \to 7$ in the previous figures.
However, Borel improved sum rum rules suppress excited states more
effectively than the Euclidean formulation presented here, as the
phenomenological side of Borel improved sum rules involves the square
of the nucleon and excitation masses.  The contribution of excited
states for a given $t_f$ is smaller for Borel improved Sum Rules and a
direct comparison of the analysis intervals must take this into
account.

   One should also acknowledge the presence of an artificial
enhancement of the lattice correlation functions at very short times.
This requires the QCD continuum model to make larger contributions
relative to the ground state.  Any structure in the lattice
correlation functions at short times reflecting structure in the
spectral density is more prevalent in this analysis.  Hence the
enhancement of the correlator gives rise to a more demanding test of
the QCD continuum model.  Structure not sufficiently accounted for by
the model is more problematic here.  The preceding results give a
strong indication that the QCD-SR-inspired continuum model provides a
sufficiently detailed description of the correlation functions in the
short time regime to allow the extraction of ground state properties.

   One might consider a similar analysis where the QCD continuum model
is replaced by a second pole as in (\ref{TwoPoles}) with two poles.
Such a comparison would demonstrate the importance of the QCD
continuum model functional form and the sensitivity of the analysis
presented here.  We consider two fits of (\ref{TwoPoles}) to the
lattice data.  Since $\xi$ was fixed in figures \ref{NmassSRT} and
\ref{NlambdaSRT} to the optimal value obtained from a four-parameter
fit to the time regime $6 \to 20$, we consider a three-parameter
two-pole fit where $\lambda_{N^*}$ is similarly fixed.  We also
present the results of a full four-parameter search for comparison.

   For the two-pole spectral density of (\ref{TwoPoles}), both the
ground state nucleon mass and coupling strength show a sensitivity to
the analysis interval.  These parameters do not form a plateau when
plotted as a function of $t_f$.  This failure of the single excitation
hypothesis for the QCD continuum model is displayed in Figure
\ref{NmassTwoPoles} for the ground state mass and Figure
\ref{NlambdaTwoPoles} for the coupling strength.

   Neither the mass nor the coupling strength extracted with a
two-pole Ansatz agree with the values obtained from the single pole
fits to the deep nonperturbative tail of the correlation function,
where evolution in Euclidean time has isolated the ground state.  A
successful continuum model must reproduce the ground state properties
obtained from the simple pole fits with $t_f \sim 20$ or 21
\cite{boundary}, as in Figures \ref{NmassSRT} and \ref{NlambdaSRT}.
For the two-pole spectral density, information on ground state
properties has been sacrificed in accommodating the deficiencies of
the spectral density in the short time regime, where the statistical
uncertainties are relatively small.  In this investigation, inclusion
of the tail of the correlation function in the analysis interval is
insufficient to determine the ground state properties when
discrepancies made by a poor continuum model need to be accommodated.

   In contrast, the QCD-Sum-Rule inspired continuum model produces
ground state properties that agree with simple pole fits to the tail
of the correlation function, and the fit parameters are stable with
respect to the analysis interval.  Indeed this analysis is very
sensitive to the continuum model Ansatz.

\section{COMPARISON WITH OTHER APPROACHES}
\label{compappr}

\subsection{Operator Product Expansions}

   Here we make a comparison of these techniques with those of other
approaches for the nucleon and $\Delta$ channels.  Since QCD Sum Rules
and the Random Instanton Liquid Model (RILM) \cite{schafer94a}
typically use the interpolator $\chi_{\rm SR}$ of (\ref{chiSR}) we
repeat the previous analysis for this interpolator and the $\Delta$
interpolator.  The OPEs for these interpolating fields are
\begin{eqnarray}
G_{\rm SR}(t) &\simeq& {3 \cdot 5 \over 2^3 \pi^4} \, \biggl ( \,
{1 \over t^6} + {2 \over 5} \, {m_q a \over t^5}
              + {1 \over 5} \, {m_q^2 a^2 \over t^4} \nonumber \\
            &&- {4 \pi^2 \over 15} \, {\bigm < : \overline q q :
                                        \bigm > a^3
                \over t^3}
              + \cdots \biggr ) \, ,
\label{OPESR}
\end{eqnarray}
and
\begin{eqnarray}
\sum_{\mu = 2,3} G_\Delta^{\mu \mu}(t) &\simeq& {3^2 \cdot 5 \over 2^5
                                                 \pi^4} \, \biggl ( \,
{1 \over t^6} + {4 \over 3} \, {m_q a \over t^5}
              + {2 \over 3} \, {m_q^2 a^2 \over t^4} \nonumber \\
            &&- {8 \pi^2 \over 3} \, {\bigm < : \overline q q :
                                        \bigm > a^3
                \over t^3}
              + \cdots \biggr ) \, ,
\label{OPEDelta}
\end{eqnarray}
where the $\mu=2,3$ Lorentz indices of the $\Delta$ interpolator have
been summed over.  Note that in this case the pole contribution to the
phenomenological side of (\ref{OPEDelta}) is $(4/3) \lambda_\Delta^2
\exp(-M_\Delta t)$.  The QCD continuum models are developed as
outlined in Section \ref{contmodel}.  A summary of the $t=6 \to 20$
fit parameters for these interpolating fields is given in Tables
\ref{Tab:chiSRC5ope} and \ref{Tab:chiDeltaC5ope}.

\subsection{Systematic Uncertainties}

   Prior to comparing with other approaches, it is important to
consider possible sources of systematic uncertainty, in addition to
the statistical uncertainties indicated in the tables.  As in most
exploratory lattice calculations, this investigation considers a
single value for the coupling constant (or lattice spacing $a$) and a
single lattice volume.  As such, extrapolations to the infinite-volume
continuum limit are not possible.

   Na\"{\i}vely, one might worry that finite volume errors are
significant in this investigation.  The lattice length in the shorter
$y$ and $z$ directions is 1.58 fm and is small relative to the
measured charge diameter, $2\bigm< r^2 \bigm>^{1/2}$, of 1.72(24) fm
for the proton \cite{simon80,dumbrajs83}.  However, it is important to
remember that the actual lattice simulations are performed with quark
masses larger than the light current quark masses of a few MeV.  For
the quark masses considered on the lattice here, the proton charge
diameter ranges from 1.0 to 1.2 fm.  For estimating finite volume
effects, the matter radius, $\left (\, (2/3)\bigm< r_u^2 \bigm> + \,
(1/3)\bigm< r_d^2 \bigm> \, \right )^{1/2}$, is a better measure.  This
diameter ranges from 1.0 fm to 1.1 fm, which is about 2/3 the length
of the shortest dimension of the lattice.

   Order $a$ lattice spacing corrections to the lattice actions can
also be a source of systematic error.  Hadron spectrum calculations
\cite{butler93} indicate these errors are about 10\% for hadron mass
ratios at $\beta=6.0$.  The main quantity of interest here is the
nucleon interpolating field coupling strength, $\lambda_N$.  Since the
nucleon mass is used to determine the lattice spacing, we have the
advantage of determining the properties of a single hadron.  Since
$M_N$ and $\lambda_N$ are linked dynamically, and $M_N$ is constrained
to the physical value, we proceed under the assumption that finite
volume and finite lattice spacing errors in $\lambda_N$ are small
relative to other sources of uncertainty.  For the $\Delta$ one should
expect systematic uncertainties from these effects to be the order of
10\%.  Quantification of these uncertainties remains a future endeavor
of lattice QCD investigations.

   The two most prominent sources of systematic uncertainty are the
quenched approximation and the extrapolation of observables to the
chiral limit.  Our present understanding of the quenched
approximation, obtained from numerical simulations of the full theory
for a few observables, is that it is an excellent approximation to the
full theory when the quark masses are sufficiently heavy.  The quark
masses considered in this investigation lie within this regime.
Hence, it remains to estimate systematic uncertainty in the linear
extrapolations of masses and coupling strengths.

   Both $M_N$ and $\lambda_N$ are well behaved in the chiral limit.
The leading nonanalytic terms of the chiral expansion are nonsingular.
Our calculations of the nucleon mass indicate the quark mass
dependence (or squared pion mass dependence) is linear to a good
approximation.  To order $m_\pi^2$, the chiral expansion for $M_N$ is
linear in $m_\pi^2$, and therefore our lattice determinations of the
nucleon mass are linearly extrapolated to the chiral limit.  Likewise,
the more phenomenological continuum threshold is taken to have a
similar mass dependence.

   To order $m_\pi^2$, the chiral expansion for $\lambda_N$ is not
linear in $m_\pi^2$.  The leading nonanalytic term is proportional to
$m_\pi^2 \log m_\pi^2$.  Here we use a phenomenological description of
the $m_\pi^2$ behavior of $\lambda_N$ to estimate the size of
systematic error in linear extrapolations.  For the interpolating
field $\chi_{\rm SR}$ it is possible to derive a simple expression for
the pion mass dependence of $\lambda_N$ \cite{lee94},
\begin{equation}
\lambda_N = \lambda_0 \left ( 1 - {3 \over 2^7 \pi^2}\, {m_\pi^2 \over
f_\pi^2} \log {m_\pi^2 \over \Lambda^2} \right ) + c \, m_\pi^2 +
\cdots \, .
\label{chiralexp}
\end{equation}
$\Lambda$ and $c$ are redundant fit parameters.  Here, $\Lambda$
is fixed to the inverse lattice spacing.

   Figure \ref{chiralext} illustrates the standard linear
extrapolation and a fit of $\lambda_0$ and $c$ in (\ref{chiralexp}) to
the lattice data.  The linear extrapolation and the result from
(\ref{chiralexp}) differ by less than one standard deviation of the
statistical uncertainty in the chiral limit.  Hence an estimate of the
combined statistical and systematic uncertainty for $\lambda_N$ may be
obtained by multiplying the statistical uncertainties in the tables by
$\sqrt{2}$.

\subsection{Comparison}

   The results summarized in Table \ref{CompNucleon} for $\lambda_{\rm
SR}$ evolved to a scale of 1.0 GeV${}^2$ in the leading log
approximation compare favorably with other approaches utilizing the
pole plus QCD continuum phenomenology \cite{DiffRenorm}.  The correct
nucleon sum rule predictions of \cite{leinweber90,SRerr} agree with
the lattice results.  Previous works addressing these issues
\cite{chu93a,chu93b,schafer94a} refer to the old and erroneous results
of Ref.\ \cite{ioffe81} and \cite{belyaev83}.  The numerous errors
contained in Ref.\ \cite{belyaev83} are discussed and corrected in
Ref.\ \cite{leinweber90}.  The corrections are significant, and
restore agreement with these lattice results and with the RILM of
\cite{schafer94a}.  The smaller value of $\lambda_{\rm SR}$ for Ref.\
\cite{chu93b} may be due to the omission of large $\alpha_s$
corrections to the Wilson coefficient of the identity operator used to
normalize $\lambda_{\rm SR}$.  Such corrections could increase their
result by approximately 20\%.

   There is some variation in the continuum threshold, $s_0$ from one
approach to another.  This is likely due to differences in the
implementation of the QCD continuum models and correlation function
enhancement in the short time regime reflecting lattice anisotropy.
This enhancement is expected to induce a larger continuum threshold,
and this is confirmed in Table \ref{CompNucleon}.

   Table \ref{CompDelta} also displays uniformity among the residue of
the pole for different approaches.  Here the corrections to Ref.\
\cite{belyaev83} are crucial \cite{leinweberPhD}.  Our analysis
displays the usual suppression of $M_\Delta$ relative to $M_N$,
typical of lattice analyses in which order $a$ and finite volume
corrections remain unaccounted \cite{butler93}.  The large value for
$M_\Delta$ from Ref.\ \cite{chu93b} is mysterious.  Once again the
value of the continuum threshold is larger than in the QCD Sum Rule
analysis as anticipated.

\section{THE PHYSICS OF INTERPOLATORS}
\label{physint}

   There has been considerable confusion surrounding the physical
significance of the baryon coupling strength $\lambda$, and more
generally the physics represented in point-to-point correlation
functions.  Here we address these issues.

   In Ref.\ \cite{schafer94a} some conclusions are drawn from an
incorrect interpretation of the physics represented in the
interpolating field coupling strength, $\lambda$.  From their study of
{\it point-to-point} correlation functions, these authors claim ``the
octet and decuplet baryons have completely different wave functions''
and that there is evidence of significant attraction in the scalar
diquark channel.  It must be stated that the phenomenological
description of $G_2(t)$ in (\ref{phenom}) involves the baryon mass,
the continuum threshold describing the effective onset of excited
states and $\lambda$, which indicates the ability of the particular
interpolator to excite the baryon from the QCD vacuum.  Information on
the nucleon wave function that might be compared with the $\Delta$
wave function is absent in point-to-point correlation functions.

   While it is tempting to compare $\lambda_N$ and $\lambda_\Delta$ in
hopes of learning something about the distribution of quarks in
baryons \cite{schafer94a}, such a comparison is not correct.  It is
important to recognize that $\lambda$ reflects properties of {\it
both} the baryon ground state {\it and} the interpolating field
itself.  Since the nucleon and $\Delta$ interpolating fields are
different it should be no surprise to discover $\lambda_N \ne
\lambda_\Delta$.  Moreover, the nucleon interpolating field is not
unique.  It is possible to construct nucleon interpolators which yield
values for $\lambda_N$ both larger and smaller than $\lambda_\Delta$.
Moreover, their anomalous dimensions are different, making any
comparison scale dependent.  In short, there is little to be learned
{}from a comparison of $\lambda_N$ and $\lambda_\Delta$.

   Moreover, $\lambda$ does not measure simply the wave function at
the origin, but rather the wave function at the origin when
\begin{enumerate}
\item the three valence quarks by themselves are in a color singlet
state;
\item the three valence quarks have the particular spin-flavor
combination demanded by the interpolator.
\end{enumerate}
The first criterion excludes the {\it majority} of the wave function
including intermediate states of simple diagrams such as a single
gluon exchange between quarks.  Similarly, the second criterion
excludes parts of the wave function that have no overlap with the
interpolating field.  These arguments also apply in general to wave
function analyses where the locality of the baryon annihilation
interpolator is relaxed and $\lambda$ is determined as a function of
quark field operator separations.  Of course, wave function analyses
also suffer from being either dependent on the gauge, or dependent on
the path of link variables selected to implement gauge invariance.

   The interpolating field acts to bias the physics represented in the
wave function.  For example, if an interpolator with predominant
scalar diquark degrees of freedom is used to annihilate the baryon,
then this part of the full baryon wave function will dominate the
extracted wave function.  The interpolating field filters out the part
of the full wave function that looks like the interpolator.  Thus,
wave function analyses allow one to learn about specific sectors of
the complete wave function.  However, it is dangerous to in turn
attribute the properties obtained from a specific sector to those of
the entire baryon wave function.

   Many have expected it would be possible to find a perfect wave
function for the creation of a hadron from the vacuum.  A perfect
smeared source would excite only the ground state hadron.  Evolution
in Euclidean time to suppress excited state contaminations would not
be required.  However, even the best smeared sources require a number
of Euclidean time steps to isolate the ground state.  This inability
to obtain a perfect smeared source has a simple explanation.  The wave
function provides information on a particular spin-flavor-color
Fock-space component of the full wave function.  Without information
on the full ground state wave function, a smeared operator will not be
orthogonal to excited states and will always generate some excited
state contaminations.

   Of course, the only way to probe the properties of hadrons is
through the use of a common interpolator or current such as the vector
current, which has overlap with most hadrons.  In this way, it is the
properties of the hadrons themselves that give rise to differences in
the extracted observables.  For example, vector current matrix
elements are determined via the calculation of three-point correlation
functions in Ref.\ \cite{leinweber91,leinweber92b,leinweber93e}, where
a variety of electromagnetic observables are reported.  These results
indicate that scalar diquark clustering in the nucleon is actually
minimal in QCD \cite{leinweber93b,sqmcomments}.

\section{CONCLUSIONS}
\label{concl}

\subsection{The QCD Continuum Model}

   The physics represented in the short time regime of point-to-point
correlation functions is described well by the QCD-Sum-Rule-inspired
continuum model.  The Laplace transform of the spectral density
appears to be sufficient to render any structure in the spectral
density insignificant in the short Euclidean time regime of
point-to-point correlators.  Similar conclusions are expected to hold
for Borel-improved sum rules.

   For the lattice correlation functions considered here, both the
identity and $m_q$ operators of the OPE are required when constructing
QCD continuum models.  Other higher order terms of the OPE make
negligible contributions to the QCD continuum model due to the
exponential suppression factor $\exp(-s_0 t)$.  In general,
consideration of the first few leading terms of a particular OPE
should be sufficient for the construction of the QCD continuum model.
The QCD continuum model is superior to the use of a second pole.

\subsection{QCD Sum Rule Test}

   The QCD continuum model effectively accounts for excited state
contributions to point-to-point correlation functions, allowing a
determination of ground state properties in the short-time regime.
However, interplay between pole and continuum model parameters leads
to rather large uncertainties.

   The smallest fit interval considered here is larger and extends
deeper into the nonperturbative regime than that considered in QCD Sum
Rule analyses.  Hence it would be inappropriate to conclude that QCD
Sum Rule techniques have been vindicated in this analysis.  However,
the success of the QCD continuum model here is encouraging.  This
analysis justifies the use of the ``crude'' QCD continuum model and
lends credence to the results of rigorous QCD Sum Rule analyses.

\subsection{Future Investigations}

   Having established that the QCD continuum model accounts for
excited state contributions and allows the determination of ground
state properties in the short-time regime, future lattice calculations
may use this technique for analyzing correlation functions that suffer
a loss of signal prior to a clear ground state domination.  In a
subsequent paper \cite{leinweber94b}, these techniques will be
exploited to investigate nucleon properties obtained from
unconventional interpolating fields.

   These techniques may be particularly helpful in the analysis of
heavy-light meson correlators.  Smeared-source to smeared-sink
correlators suffer from large statistical uncertainties, while
point-to-point correlators suffer from a loss of signal.  One of these
two correlators are needed in order to determine the meson decay
constant.  In addition, the mass splittings are relatively small in
these systems, making isolation of the ground state by Euclidean time
evolution inefficient.  On the other hand, smeared-local correlators
are well behaved and allow a clean determination of the ground state
mass.  Inclusion of the QCD continuum model in the phenomenological
description of the correlator allows the analysis of the
point-to-point correlator and provides a mechanism to account for the
problematic excited state contaminations.  With the pole position
previously determined, a reliable extraction of the meson decay
constant may be possible.

\acknowledgements

   The correlation functions used in this analysis were obtained in
collaboration with Terry Draper and Richard Woloshyn in Ref.\
\cite{leinweber91}.  I would like to thank J. Pasupathy for
rekindling my interest in these issues and for his input during the
early stages of this research.  I also thank Thomas Cohen, Dick
Furnstahl, Robert Perry and Greg Kilcup for illuminating discussions.
Thanks also to the Institute for Nuclear Theory for their hospitality
during the summer session workshop on ``Phenomenology and Lattice
QCD'' where some of the ideas presented here were developed.  This
research is supported in part by the National Science Foundation under
grants PHY-9203145, PHY-9258270, PHY-9207889 and PHY-9102922 and by
the Department of Energy under grant DE-FG02-93ER-40762.


\begin{thebibliography}{10}

\bibitem[*]{presaddr} Present address: TRIUMF, 4004 Wesbrook Mall,
Vancouver, BC, V6T 2A3, Canada.

\bibitem{shifman79}
M.~A. Shifman, A.~I. Vainshtein, and Z.~I. Zakharov, {\it Nucl.\
Phys.\ } {\bf B147},  385, 448  (1979).

\bibitem{leinweber90}
D.~B. Leinweber, {\it Ann.\ Phys.\ }(N.Y.) {\bf 198},  203  (1990).

\bibitem{leinweber94e} D.~B. Leinweber, in {\em Lattice '93, {\rm
Proceedings of the International Symposium}}, Dallas, TX, 1993, edited
by T. Draper, S. Gottlieb, A. Soni, and D. Toussaint ({\it Nucl.\
Phys.\ B} ({\it Proc.\ Suppl.\ }), {\bf 34}, (1994) 407).

\bibitem{chu93a} M. Chu, J. Grandy, S. Huang, and J. Negele, {\it
Phys.\ Rev.\ Lett.\ } {\bf 70}, 255 (1993).

\bibitem{chu93b} M. Chu, J. Grandy, S. Huang, and J. Negele, {\it
Phys.\ Rev.\ D} {\bf 48}, 3340 (1993).

\bibitem{leinweber94b} D.~B. Leinweber, \protect Nucleon properties
from unconventional interpolating fields. OSU-PP \#94--0333
(unpublished).

\bibitem{sakurai67} J. Sakurai, {\em Advanced Quantum Mechanics}
(Addison-Wesley, Reading, 1967).

\bibitem{ioffe81}
B.~L. Ioffe, {\it Nucl.\ Phys.\ } {\bf B188},  317  (1981).

\bibitem{ioffe83}
B.~L. Ioffe, {\it Z. Phys. C} {\bf 18},  67  (1983).

\bibitem{leinweber91} D.~B. Leinweber, R.~M. Woloshyn, and T. Draper,
{\it Phys.\ Rev.\ D} {\bf 43}, 1659 (1991).

\bibitem{cabibbo82}
N. Cabibbo and E. Marinari, {\it Phys.\ Rev.\ Lett.\ } {\bf 119B},  387
  (1982).

\bibitem{correlations} This proved sufficient to achieve good
statistical independence of the ensemble. An examination of a number
of observables, configuration by configuration, revealed no apparent
long-range correlation in our gauge configuration sample.

\bibitem{efron79}
B. Efron, {\it SIAM Rev.\ } {\bf 21},  460  (1979).

\bibitem{michael94}
C. Michael, {\it Phys.\ Rev.\ D} {\bf 49},  2616  (1994).

\bibitem{rpp92} {Particle Data Group, K. Hikasa, {\it et al.\ }}, {\it
Phys.\ Rev.\ D} {\bf 45}, III.38 (1992).

\bibitem{lepage92} G.~P. Lepage, in {\em Lattice '91, {\rm Proceedings
of the International Symposium}}, Tsukuba, Japan, 1991, edited by
M. Fukugita, Y. Iwasaki, M.  Okawa, and A. Ukawa ({\it Nucl.\ Phys.\
B} ({\it Proc.\ Suppl.\ }), {\bf 26}, 45, (1992)).

\bibitem{lepage93} G.~P. Lepage and P.~B. Mackenzie, {\it Phys.\ Rev.\
D} {\bf 48}, 2250 (1993).

\bibitem{richards87} D. Richards, C. Sachrajda, and C. Scott, {\it
Nucl.\ Phys.\ } {\bf B286}, 683 (1987).

\bibitem{daniel92} D. Daniel {\it et~al.}, {\it Phys.\ Rev.\ D} {\bf
46}, 3130 (1992).

\bibitem{mfimq} The mean-field improved suggestion of $m_q \to 8 \,
\kappa_{cr} \, m_q$ was also examined. However, the following analysis
is not sensitive to the precise definition of the quark mass.

\bibitem{leinweber94c}
D.~B. Leinweber, \protect research in progress. (unpublished).

\bibitem{anomdim} For clarity, the anomalous dimension scaling factors
for the operators appearing in the QCD continuum model have been
suppressed in the derivation of Section \protect\ref{contmodel}. While
their inclusion in the OPE of QCD-SR analyses is essential, their
inclusion in the QCD continuum model presented here is not. This has
been verified by including the anomalous dimension scaling factors,
and the changes in the fit parameters are small relative to the
statistical uncertainties.

\bibitem{boundary} For time slices $t > 21$, lattice boundary
artifacts may become significant.

\bibitem{schafer94a} T. {Sch\"afer}, E. Shuryak, and J. Verbaarschot,
{\it Nucl.\ Phys.\ } {\bf B412}, 143 (1994).

\bibitem{simon80} G.~G. Simon, {\it Z.\ Naturforsch} {\bf 35a}, 1
(1980).

\bibitem{dumbrajs83}
O. Dumbrajs, {\it Nucl.\ Phys.\ } {\bf B216},  277  (1983).

\bibitem{butler93} F. Butler {\it et~al.}, {\it Phys.\ Rev.\ Lett.\ }
{\bf 70}, 2849 (1993).

\bibitem{lee94} S.~H. Lee, S. Choe, T.~D. Cohen, and D.~K. Griegel,
\protect QCD Sum Rules and Chiral Logarithms. U. MD Preprint
94-???. (unpublished).

\bibitem{DiffRenorm} The values reported here differ from that
presented in Ref.\ \protect\cite{leinweber93c} due to minor
differences in the renormalization procedure.

\bibitem{SRerr} The uncertainties in the QCD Sum Rule results reflect
the spread of values obtained from the consideration of all the
nucleon sum rules in Ref.\ \protect\cite{leinweber90} involving
$\chi_{\rm SR}$.

\bibitem{belyaev83} V.~M. Belyaev and B.~L. Ioffe, {\it Sov.\ Phys.\
JETP} {\bf 57}, 716 (1983).

\bibitem{leinweberPhD} D.~B. Leinweber, Ph.D. thesis, McMaster
University, Hamilton, ON, Canada, 1988.

\bibitem{leinweber92b} D.~B. Leinweber, T. Draper, and R.~M. Woloshyn,
{\it Phys.\ Rev.\ D} {\bf 46}, 3067 (1992).

\bibitem{leinweber93e} D.~B. Leinweber, T. Draper, and R.~M. Woloshyn,
{\it Phys.\ Rev.\ D} {\bf 48}, 2230 (1993).

\bibitem{leinweber93b} D.~B. Leinweber, {\it Phys.\ Rev.\ D} {\bf 47},
5096 (1993).

\bibitem{sqmcomments} While some have objected \cite{schafer94a} that
these properties of QCD are not in accord with the simple quark model
\protect\cite{isgur78,isgur80}, it is important to remember that quark
models are not QCD while lattice calculations are. One must expect
differences in their predictions. In particular, charge radii are
sensitive to long distance nonperturbative physics. It should not be
too surprising to find predictions based on a single phenomenological
gluon exchange to fail.

\bibitem{leinweber93c} D.~B. Leinweber, in {\em Lattice '93, {\rm
Proceedings of the International Symposium}}, Dallas, TX, 1993, edited
by T. Draper, S. Gottlieb, A. Soni, and D. Toussaint ({\it Nucl.\
Phys.\ B} ({\it Proc.\ Suppl.\ }), {\bf 34}, (1994) 407).

\bibitem{isgur78} N. Isgur, G. Karl, and R. Koniuk, {\it Phys.\ Rev.\
Lett.\ } {\bf 41}, 1269 (1978), \protect {\it ibid.} {\bf 45}, 1738
(E) (1980).

\bibitem{isgur80}
N. Isgur and G. Karl, {\it Phys.\ Rev.\ D} {\bf 21},  3175  (1980).

\end{thebibliography}

\mediumtext
\begin{table}
\caption{$\bigm < \chi_1 \overline\chi_1 \bigm >$:
         Four-parameter search for the pole plus QCD continuum model.}
\label{Tab:chi11C5ope}
\setdec 0.000000
\begin{tabular}{lcccc}
Parameter  &$\kappa_1=0.152$ &$\kappa_2=0.154$ &$\kappa_3=0.156$
           &$\kappa_{\rm cr}=0.159\,8(2)$\\
\tableline
$M_N a$ &\dec 1.109(8) &\dec 0.983(8) &\dec 0.858(8) &\dec
0.628(17)\tablenotemark[1] \\
$\lambda_1 a^3$ $(\times 10^{-2})$
 &\dec 1.17(5)    &\dec 0.94(4)   &\dec 0.75(3)   &\dec 0.38(7)  \\
$s_0 a$
 &\dec 1.68(3)    &\dec 1.58(3)   &\dec 1.49(4)   &\dec 1.32(7)  \\
$\xi$
 &\dec 6.83(10)   &\dec 6.74(9)   &\dec 6.62(9)   &\dec 6.42(19) \\
$\xi$ from OPE fit
 &\dec 5.3(1)     &\dec 5.6(1)    &\dec 5.8(1)    & \\
\end{tabular}
\tablenotetext[1]{The physical proton mass sets the lattice spacing
$a=0.132(4)$ fm.}
\end{table}

\begin{table}
\caption{${1 \over 4}
          \bigm < \chi_{\rm SR} \overline\chi_{\rm SR} \bigm >$:
         Four-parameter search for the pole plus QCD continuum model.}
\label{Tab:chiSRC5ope}
\setdec 0.000000
\begin{tabular}{lcccc}
Parameter  &$\kappa_1=0.152$ &$\kappa_2=0.154$ &$\kappa_3=0.156$
           &$\kappa_{\rm cr}=0.159\,8(2)$\\
\tableline
$M_N a$
 &\dec 1.106(8)   &\dec 0.980(9)  &\dec 0.863(9)  &\dec 0.639(18) \\
$\lambda_1 a^3$ $(\times 10^{-2})$
 &\dec 1.14(5)    &\dec 0.92(4)   &\dec 0.76(3)   &\dec 0.42(8)  \\
$s_0 a$
 &\dec 1.63(3)    &\dec 1.53(3)   &\dec 1.45(4)   &\dec 1.29(7)  \\
$\xi$
 &\dec 5.46(7)    &\dec 5.25(7)   &\dec 5.02(7)   &\dec 4.61(14) \\
$\xi$ from OPE fit
 &\dec 4.5(5)     &\dec 4.6(4)    &\dec 4.6(3)   & \\
\end{tabular}
\end{table}

\begin{table}
\caption{$\bigm < \chi_\Delta \overline\chi_\Delta \bigm >$:
         Four-parameter search for the pole plus QCD continuum model.}
\label{Tab:chiDeltaC5ope}
\setdec 0.000000
\begin{tabular}{lcccc}
Parameter  &$\kappa_1=0.152$ &$\kappa_2=0.154$ &$\kappa_3=0.156$
           &$\kappa_{\rm cr}=0.159\,8(2)$\\
\tableline
$M_\Delta a$
 &\dec 1.164(8)    &\dec 1.054(8)  &\dec 0.953(9)  &\dec 0.758(18)  \\
$\lambda_\Delta a^3$ $(\times 10^{-2})$
 &\dec 2.91(11)    &\dec 2.32(9)   &\dec 1.82(8)   &\dec 0.82(17)  \\
$s_0 a$
 &\dec 1.84(2)     &\dec 1.78(2)   &\dec 1.73(3)   &\dec 1.62(5)  \\
$\xi$
 &\dec 10.3(1)     &\dec 10.1(1)   &\dec 10.0(1)   &\dec 9.8(3) \\
$\xi$ from OPE fit
 &\dec 7.0$\pm1.5$ &\dec 7.3$\pm1.3$ &\dec 7.6$\pm1.2$  & \\
\end{tabular}
\end{table}

\begin{table}
\caption{Comparison with selected results for the nucleon channel
$\bigm < \chi_{\rm SR} \overline\chi_{\rm SR} \bigm >$.}
\label{CompNucleon}
\setdec 0.000(0)
\begin{tabular}{llccc}
Approach &Ref.      &$M_N$ &$\lambda_{\rm SR}$(1 GeV${}^2$) &$s_0$ \\
         &          &(GeV) &(GeV${}^3$)        &(GeV) \\
\tableline
This work &
 &\dec 0.96(3)      &\dec 0.027(5)  &\dec 1.92(11) \\
QCD Sum Rules       &Leinweber \cite{leinweber90}
 &\dec 1.06(18)     &\dec 0.031(6)  &\dec 1.69(15)  \\
Lattice ($x$-space)  &Chu {\it et al.} \cite{chu93b}
 &\dec 0.95(5)      &\dec 0.022(4)  &\dec $<$1.4  \\
Instanton Liquid    &Sch\"afer, {\it et al.} \cite{schafer94a}
 &\dec 0.96(3)      &\dec 0.032(1)  &\dec 1.92(5)  \\
\end{tabular}
\end{table}

\begin{table}
\caption{Comparison with selected results for the $\Delta$
resonance channel
$\bigm < \chi_\Delta \overline\chi_\Delta \bigm >$.}
\label{CompDelta}
\setdec 00.000(00)
\begin{tabular}{llccc}
Approach &Reference &$M_\Delta$
&$\lambda_\Delta$(1 GeV${}^2$) &$s_0$ \\
         &          &(GeV)
&(GeV${}^3$)                   &(GeV) \\
\tableline
This work &
 &\dec 1.13(4)      &\dec 0.024(6)  &\dec 2.39(10) \\
QCD Sum Rules &Leinweber \cite{leinweberPhD}
 &\dec 1.36         &\dec 0.030     &\dec 1.58 \\
Lattice ($x$-space) &Chu {\it et al.} \cite{chu93b}
 &\dec 1.43(8)      &\dec 0.037(6)  &\dec 3.21(34)  \\
Instanton Liquid     &Sch\"afer, {\it et al.} \cite{schafer94a}
 &\dec 1.44(7)      &\dec 0.033(5)  &\dec 1.96(10)  \\
\end{tabular}
\end{table}

%
%
\narrowtext

\begin{figure}
\caption{
A fit to the lattice correlation function (bullets) from $t=7 \to 20$
utilizing a two pole Ansatz for the spectral density (solid curve).
Individual pole contributions are illustrated by the dashed lines.
When time slice 6 is included in the fit, the mass of the second pole
increases to accommodate the earlier time slice.}
\label{twopolefit}
\end{figure}

\begin{figure}
\caption{
Comparison of the fit of the $1/t^6$ term with the ultraviolet cutoff
correction (dashed line) of (\protect\ref{uvcutoff}) and the
corresponding curve for $1/t^6$ alone (solid line).  The lattice
cutoff correction term is negligible by $t-t_0=2$.  In the following,
the source and first point of the correlator are discarded and all
fits begin at $t=6$. }
\label{cutoffcomp}
\end{figure}

\begin{figure}
\caption{QCD continuum model contributions.  From top down the lines
represent the total continuum model contribution, followed by the
individual contributions of the identity operator, $m_q$ correction,
quark condensate and finally the $m_q^2$ correction. }
\label{ContModelCont}
\end{figure}

\begin{figure}
\caption{ OPE contributions.  From top down, at time slice 6, the
lines represent the sum total of the first four terms of the OPE,
followed by the individual contributions of the identity operator,
$m_q$ correction, quark condensate and finally the $m_q^2$ term.
Line types are as in Figure \protect\ref{ContModelCont}.  The
crossing of the quark condensate contribution with the identity
operator at $t=8$ indicates the breakdown of the truncated OPE.  The
shaded areas indicate the regions typically excluded from
Borel-transformed QCD Sum Rule analyses.}
\label{OPEcont}
\end{figure}

\begin{figure}
\caption{
The two-point correlator at $\kappa=0.152$ for the nucleon
interpolating field $\chi_1$ of (\protect\ref{chiN1}).  The fits for
the 12 analysis intervals are illustrated.  The source position is at
$t_0 = 4$.  Neither the source nor $t = 5$ are included in the fit as
indicated in the discussion surrounding Figure
\protect\ref{cutoffcomp}. }
\label{NuclCorrFn}
\end{figure}

\begin{figure}
\caption{
The nucleon mass determined in each analysis interval plotted as a
function of $t_f$.  In this and the following figure, bullets
correspond to pole plus QCD continuum fits from $t = 6 \to t_f$, and
open squares, offset for clarity, illustrate a simple pole fit to the
region $t_f - 7 \to t_f$ which was selected to give similar
statistical uncertainties in the nucleon mass at $t=20$. }
\label{Nmass}
\end{figure}

\begin{figure}
\caption{
The nucleon coupling strength determined in each analysis interval
plotted as a function of $t_f$.  Symbols are as in Figure
\protect\ref{Nmass}. }
\label{Nlambda}
\end{figure}

\begin{figure}
\caption{
The quark mass dependence of $\xi$ for mean-field-improved wave
function renormalization (solid curve) and na\"{\i}ve wave function
normalization (dashed curve). }
\label{xiMassDep}
\end{figure}

\begin{figure}
\caption{Correlation function fits for each of the 15 analysis
intervals $t=6 \to t_f$, where $t_f=9 \to 23$ for the QCD-SR method
test. }
\label{NcorrelSRT}
\end{figure}

\begin{figure}
\caption{
The nucleon mass determined in each analysis interval plotted as a
function of $t_f$.  In this and the following figure, bullets
correspond to pole plus QCD continuum fits from $t = 6 \to t_f$, and
open squares illustrate a simple pole fit to the region $t_f - 7 \to
t_f$ as in Figure \protect\ref{Nmass}. }
\label{NmassSRT}
\end{figure}

\begin{figure}
\caption{
The nucleon coupling strength determined in each analysis interval
plotted as a function of $t_f$.  Symbols are as in Figure
\protect\ref{NmassSRT}. }
\label{NlambdaSRT}
\end{figure}

\begin{figure}
\caption{ The nucleon ground state mass determined in each analysis
interval, $t=6 \to t_f$, plotted as a function of $t_f$ for a spectral
density involving two poles, as opposed to a pole plus QCD continuum
model.  Bullets correspond to a three-parameter fit with
$\lambda_{N^*}$ fixed as described in the text.  Open circles
illustrate the full four-parameter fit.  The extracted mass depends on
the analysis interval and fails to agree with the simple
pole fits (open squares) from $t_f - 7 \to t_f$ at $t_f \sim 20$ or
21.  This confirms the sensitivity of this analysis to the QCD
continuum model Ansatz.}
\label{NmassTwoPoles}
\end{figure}

\begin{figure}
\caption{ The nucleon coupling strength plotted as a function of $t_f$
for a spectral density involving two poles.  Symbols are as in Fig.\
\ref{NmassTwoPoles}.  The dependence on the analysis interval and
failure to agree with the simple pole fits (open squares) at $t_f \sim
20$ or 21 confirms the sensitivity of this analysis.}
\label{NlambdaTwoPoles}
\end{figure}

\begin{figure}
\caption{ Extrapolation of $\lambda_{\rm SR}$ to the chiral limit.
The solid line indicates the standard linear extrapolation, while the
dashed curve illustrates the two-parameter fit of
(\protect\ref{chiralexp}) to the lattice data.}
\label{chiralext}
\end{figure}

\end{document}